# Accidental degeneracy of double Dirac cones in a phononic crystal

Ze-Guo Chen,[1] Xu Ni,[1] Ying Wu,[2] Cheng He,[1] Xiao-Chen Sun,[1] Li-Yang Zheng,[1] Ming-Hui Lu,[1,*] Yan-Feng Chen[1]

[1] National Laboratory of Solid State Microstructures & Department of Materials Science and Engineering, Nanjing University, Nanjing 210093, People's Republic of China

[2] Division of Mathematical and Computer Sciences and Engineering, King Abdullah University of Science and Technology (KAUST), Thuwal 23955-6900, Saudi Arabia

Artificial honeycomb lattices with Dirac cone dispersion provide a macroscopic platform to study the massless Dirac quasiparticles and their novel geometric phases. In this paper, a quadruple-degenerate state is achieved at the center of Brillouin zone (BZ) in a two-dimensional honeycomb lattice phononic crystal, which is a result of accidental degeneracy of two double-degenerate states. In the vicinity of the quadruple-degenerate state, the dispersion relation is linear. Such quadruple degeneracy is analyzed by rigorous representation theory of groups. Using $\vec{k}\cdot\vec{p}$ method, a reduced Hamiltonian is obtained to describe the linear Dirac dispersion relations of such quadruple-degenerate state, which is well consistent with the simulation results. Near such accidental degeneracy, we observe some unique wave propagating properties, such as defect insensitive propagating character and Talbot effect.

PACS numbers: 43.20. +g, 73.22.Pr, 43.35. +d

[*] e-mail: luminghui@nju.edu.cn

## I. INTRODUCTION

Many unique phenomena in graphene such as quantum Hall effect, *Zitterbewegung,* Klein paradox and pseudo-diffusion, are attributed to the unique dispersion relation of massless quasiparticles solved by Dirac equation.[1-5] The eigen-energy $E$ is linearly proportional to the wave vector $k$ at the six corners of the hexagonal boundary of the BZ. The upper and lower bands near the $K$ point act as two cones touching at one degenerate point, which is the so-called Dirac point and such conical dispersion is called Dirac cone. Compared to graphene or photonics and phononics with Dirac cone dispersion at the corner of the BZ,[6] the recent observation of Dirac cones at the center of BZ in photonic and phononic crystals has also attracted much attention. Under certain circumstances, those Dirac cones can be mapped into a zero-index material, whose parameters (e.g. permittivity and permeability in electromagnetics, effective mass density and reciprocal of bulk modulus in acoustics) are both vanishing.[7-11] It provides a new method to achieve zero-index materials with simple photonic and phononic crystals so that many interesting properties such as wave shaping and cloaking are easily demonstrated.[8, 9][12, 13]

Linear dispersion is a key feature of a Dirac cone. However, the linear dispersion at a finite frequency is in general forbidden at the center of the BZ, because of time-reversal symmetry.[14, 15] The previously mentioned linear dispersion relations at the center of the BZ in classical wave systems are achieved by *accidental degeneracy*. In 2D photonic and phononic crystals with $C_{4v}$ symmetry,[7-9] it has been demonstrated that the accidental degeneracy of a monopolar state (or a quadrupolar state) and a double-degenerate dipolar state can lead to three folded degeneracy showing linear dispersion in the vicinity of the $\Gamma$ point. In addition to the linear bands, there is a flat band intersect with them at the Dirac point. This is a major difference from the Dirac cones observed in graphene system, in which only two linear bands touch at the Dirac point. From a perturbation theory, it has been demonstrated that the Dirac cone induced by the triple-degenerated states at the center of the BZ is not a truly a Dirac cone because the reduced Hamiltonian cannot be casted into a Dirac equation (corresponding to three states) and the Berry phase equals to zero. Therefore, to be more precise, it is called a Dirac-like cone.[16] Recently, the double Dirac cone degeneracy at the BZ center has been predicted in triangular-lattice metamaterials with $C_{6v}$ symmetry.[17] However, to the best of our knowledge, such quadruple generate linear dispersion is still not materialized in normal dielectric photonic crystals or phononic crystals, and furthermore, the underlying physics, such as the reduced Hamiltonian and the Berry phase, still remains unexplored. Meanwhile, such quadruple-degenerate Dirac-like cone states may also be expected rich physics that give rise to unique wave propagating properties to be explored.

In this paper, we demonstrate that a quadruple-degenerate state can be created at the BZ center by accidental degeneracy of $E_1$ and $E_2$ modes in a two-dimensional phononic crystal with honeycomb lattice. In the vicinity of the quadruple-degenerate state, the dispersion relation is linear, forming four cones touching at their vertices. Different than the Dirac-like cone induced by triple-degenerate state, there is no flat branch. We perform a symmetry analysis to show the dispersion relation is linear and employ a $\vec{k} \cdot \vec{p}$ method to accurately predict the slope of the linear dispersion. The results of the $\vec{k} \cdot \vec{p}$ method also unambiguously reveal that the reduced Hamiltonian can be mapped into 4×4 massless Dirac equation but the Berry phase cancels out due to the absence of imaginary part in Dirac equation. Moreover, based on such quadruple Dirac degeneracy, a novel defect insensitive propagating and Talbot effects in such phononic crystals are well described with the finite element simulation.

## II. MODELS AND METHODS

### A. Physical system

The 2D PC considered here is composed of a honeycomb array of iron cylinders embedded in water ( $\rho_1 = 1000 kg/m^3, c_1 = 1490 m/s$ ,and $\rho_2 = 7670 kg/m^3, c_2 = 6010 m/s$ , where $\rho$ and $c$ denote mass density and velocity of sound and subscripts 1 and 2 correspond to water and iron, respectively). The distance between two cylinders in one unit cell is $d = 1m$, the lattice constant is $a = \sqrt{3}m$ and the radius of the cylinder is $r = 0.3710m$. Because of the large difference of velocities between iron and water, the shear modes inside the iron cylinders can be ignored.[18, 19]

### B. Band structure and Degenerate Bloch states

Figure 1(a) shows the band structure of the PC. It exhibits four bands touching linearly at one point at the frequency $\omega_0 = 892.77 Hz$ at the center of the BZ, forming four cones. Such double Dirac cones are resulted from accidental degeneracy, which is clearly demonstrated in Fig. 1(b) when the radii of the cylinders are changed into $r = 0.32m$. The quadruple-degenerate state shown in Fig. 1(a) is splited into two double-degenerate states and the linear dispersion disappears. Since we are interested in the linear dispersion near the $\Gamma$ point, we choose a region denoted by the red rectangle shown in Fig. 1 as our focus. Different from triply degenerate case,[7] there is no flat branch intersection in our model. Four cones are formed by the linear branches and touches at one point at a frequency of $\omega_0 = 892.77 Hz$ with tolerance $10^{-6}$.These four eigen degenerate states are shown in Figs. 2(a-d). All numerical simulations are accomplished by adopting the COMSOL Multiphysics, a commercial package based on the finite-element method.

### C. Symmetry analysis

By examining the symmetry of the eigenstates at the degenerate point, one can conclude if the dispersion near that point is linear or not.[20] The eigenstates of the quadruple-degenerate state are plotted in Figs. 2(a)-2(d). According to group theory, the Bloch states at $\Gamma$ point with $C_{6v}$ symmetry can be described as the basis of the irreducible representation based on the symmetry properties of the states.[21] The four eigen states shown in Figs. 2(a)-2(d) match well with the four Bloch basis functions as shown in Table 1. The two double-degenerated states that result in the quadruple-degenerate state when they meet together corresponding to $E_1$ and $E_2$ irreducible representations respectively. When any symmetry operation of $C_{6v}$ is performed, the eigenfunction of $E_1$ state transforms like $x$ and $y$, and $E_2$ state transforms like $2xy$ and $x^2 - y^2$. The wave functions near the degenerated point can be expressed as linear combinations of the degenerate states. (These linear combination wave functions are also adopted in $\vec{k} \cdot \vec{p}$ method in next section.) It should be noted that even under the same operation, the transformation of the Bloch state constructed of $E_1$ is different from that of $E_2$. Under the symmetry operations of $\sigma_x$ and $\sigma_y$, $E_1$ and $E_2$ representation can further be classified into four states. Consequently, the double Dirac cones induced by the accidental degeneracy shown in Fig. 1(a) is supported by group theory analysis.[17] The states near the Dirac point is linear combinations of the Bloch states shown in Fig. 2(a)-2(d) with parity of $\sigma_x$ or $\sigma_y$, and the wave equation should possess the same parity. Moreover, considering the compatibility relation along $\Gamma K$ and $\Gamma M$ directons, $E_1 = A + B, E_2 = A + B$, where A is full symmetry representation which indicates the existence of isotropic linear dispersion.[21]

## III. RESULTS AND DISCUSSION

### A. The reduced Hamiltonian and slopes of linear dispersion

Here, we resort to the well-known $\vec{k}\cdot\vec{p}$ method in electronics to analyze our phononic model.[16] We can rewrite the Bloch functions near $\vec{k}_0$ as linear combinations of four $\vec{k}_0$ states at $\vec{k}_0$. Substituting such function into wave equation with periodic boundary conditions, we can get[16]

$$\det\left|H - \frac{\omega_{n\vec{k}}^2 - \omega_{j0}^2}{c_1^2}I\right| = 0,\qquad(1)$$

where $n$ denotes the band index, $\vec{k}$ is the Bloch wave vector, and $H$ is the reduced Hamiltonian matrix with element $H_{ij} = i\vec{k}\cdot\vec{L}_{ij}$, $i$ and $j$ are subscripts of matrix elements. Here, $\vec{L}_{ij}$ is a real vector in $xy$-plane. The $x$ component of $\vec{L}_{ij}$ can be numerically calculated from the Bloch states as,

$$L_{ij}(x) = \frac{(2\pi)^2}{\Omega}\left(\int_{unitcell}\psi_{i\vec{k}_0}^*(\vec{r})\times\frac{2\frac{\partial\psi_{j\vec{k}_0}^*(\vec{r})}{\partial x}}{\rho_r(\vec{r})}d\vec{r} + \oint\psi_{i\vec{k}_0}^*(\vec{r}_0)\times\frac{\rho_r-1}{\rho_r}\times\cos(\theta)\times\psi_{j\vec{k}_0}(\vec{r}_0)d\vec{r}_0\right).$$

(2)

where $\rho_r(\vec{r}) = \rho(\vec{r})/\rho_1$. The $y$ component of $\vec{L}_{ij}$ can be calculated using the same process. In Eq. (2), only eight vectors are nonzero. Considering the anti-symmetrical property $\vec{L}_{ij} = -\vec{L}_{ji}$, only four vectors are independent, and the relationships of these four vectors can be described as [shown Fig. 3(a)]

$$\vec{L}_{13} = -\vec{L}_{24}, \vec{L}_{23} = \vec{L}_{14}, \vec{L}_{13}\cdot\vec{L}_{23} = 0.\qquad(3)$$

$$\begin{aligned}\vec{L}_{13} &= (-0.03768, 4.0005), \vec{L}_{14} = (4.0009, 0.03768)\\ \vec{L}_{23} &= (4.0011, 0.03766), \vec{L}_{24} = (0.03766, -0.0014)\end{aligned}.$$

Thus, the reduced Hamiltonian $H$ can be casted into:

$$H = \begin{vmatrix} 0 & 0 & i\vec{k}\cdot\vec{L}_{13} & i\vec{k}\cdot\vec{L}_{14} \\ 0 & 0 & i\vec{k}\cdot\vec{L}_{23} & i\vec{k}\cdot\vec{L}_{24} \\ -i\vec{k}\cdot\vec{L}_{13} & -i\vec{k}\cdot\vec{L}_{23} & 0 & 0 \\ -i\vec{k}\cdot\vec{L}_{14} & -i\vec{k}\cdot\vec{L}_{24} & 0 & 0 \end{vmatrix} \quad (4)$$

For the Bloch state at $\vec{k}$ in the vicinity of the $\Gamma$ point, the angle between the Bloch wave vector and $\vec{L}_{13}$ is $\theta$, by using Eq. (1), we can get dispersion relations of two Dirac cones

$$\frac{\Delta\omega}{\Delta k} = \pm\frac{|\vec{L}|c_1^2}{2\omega_0} \quad (5)$$

Where $\Delta\omega = \omega - \omega_0$ and $\Delta k = k - k_0$, Eq. (5) is linear in $\Delta k$ and is independent of $\theta$ indicating the dispersion relation is isotropic, which could be confirmed by the numerical simulations shown by solid dots in [Fig. 3(b)], and the isotropic equi-frequency contours (EFCs) shown in Figs. 4(b1) and (b2) results from the coupling of the degenerate Bloch state,[17] which matches well with the prediction of the group theory.

Knowing the length of $\vec{L}_{13}$, we can analytically calculate the dispersion relations from Eq. (5) as the red lines shown in Fig. 3(b), which overlaps with the solid dots well in the BZ center. It should be noted that although Eq. (5) exhibits only two roots, there should be four solutions to Eq (1), which means each root represented in Eq. (5) corresponds to two identical (degenerate) solutions. This is an important result, as it indicating rather than having Dirac cones with different linear slopes, the Dirac cones produced here by quadruple-degenerate state have identical slopes. This theoretical prediction is consistent with the simulated band structure [Fig. 3(b)]. The EFCs of these four bands are plotted in Fig. 4. Near the Dirac point, there is only one circles in the EFCs, verifying the isotropic property [Figs. 4(b1) and 4(b2) are identical]. Away from the Dirac point, apparently seen from Figs. 4(a) and 4(c), the EFCs for different bands are different and their hexagonal shapes indicate the dispersion is anisotropic.

## B. The Berry phase

The linear dispersion at the $\Gamma$ point described above looks very similar to the Dirac point at the BZ corner studied earlier.[5] It has been reported that in a phononic crystal the Dirac point at the corner of the BZ carries nonzero Berry phase,[22] while the Dirac-like point with triple degeneracy at $\Gamma$ point carries zero Berry phase.[16] Now, we have achieved double Dirac cone at the $\Gamma$ point by quadruple-degenerate state, does it carry zero or nonzero Berry phase? To answer this question, we perform the following analysis.

The eigenfunction of the $H$ near $\vec{k}_0$ is

$$\varphi_1(\vec{k}) = \frac{1}{\sqrt{2}} \begin{pmatrix} i\sin\theta \\ -i\cos\theta \\ 0 \\ 1 \end{pmatrix} e^{i\vec{k}\cdot\vec{r}} \quad \varphi_2(\vec{k}) = \frac{1}{\sqrt{2}} \begin{pmatrix} i\cos\theta \\ i\sin\theta \\ 1 \\ 0 \end{pmatrix} e^{i\vec{k}\cdot\vec{r}}$$

$$\varphi_3(\vec{k}) = \frac{1}{\sqrt{2}} \begin{pmatrix} -i\sin\theta \\ i\cos\theta \\ 0 \\ 1 \end{pmatrix} e^{i\vec{k}\cdot\vec{r}} \quad \varphi_4(\vec{k}) = \frac{1}{\sqrt{2}} \begin{pmatrix} -i\cos\theta \\ -i\sin\theta \\ 1 \\ 0 \end{pmatrix} e^{i\vec{k}\cdot\vec{r}}$$

(6)

where $\theta$ is the angle between $\vec{k}$ and $\vec{L}_{13}$. We can calculate the Berry phase as,

$$\Gamma_i = i \cdot \oint \langle \varphi_i(\vec{k}) | \nabla_{\vec{k}} | \varphi_i(\vec{k}) \rangle \cdot d\vec{k}. \tag{7}$$

Taking $\varphi_1(\vec{k})$ as an example, we can write $\sin\theta = \dfrac{uk_x - vk_y}{k}$, $\cos\theta = \dfrac{vk_x + uk_y}{k}$, where $u^2 + v^2 = 1$. Substituting $\varphi_1(\vec{k})$ into Eq. (7), we find

$$\begin{aligned}
\Gamma_i &= i \cdot \oint \langle \varphi_i(\vec{k}) | \nabla_{\vec{k}} | \varphi_i(\vec{k}) \rangle \cdot d\vec{k} \\
&= \frac{i}{2} \oint [-\frac{i}{k}(uk_x - vk_y), \frac{i}{k}(vk_x + uk_y), 0, 1] \\
&\quad \times \left\{ \begin{bmatrix} \frac{i}{k}(u\vec{x} - v\vec{y}) \\ \frac{-i}{k}(v\vec{x} + u\vec{y}) \\ 0 \\ 0 \end{bmatrix} + \begin{bmatrix} \frac{i}{k}(uk_x - vk_y) \\ \frac{-i}{k}(vk_x + uk_y) \\ 0 \\ 1 \end{bmatrix} i\vec{r} \right\} \cdot d\vec{k}, \quad (8) \\
&= \frac{i}{2} \oint [\frac{1}{k^2}(k_x dk_x + k_y dk_y)] + \frac{i}{2} \oint 2i\vec{r} d\vec{k} \\
&= 0
\end{aligned}$$

The same result can be obtained if we use any $\varphi_i(\vec{k})$ in Eq. (7), which means the Berry phase for our system at $\Gamma$ point is zero.

In other words, the $H$ of our system can be written as

$$H = -\vec{L}_{14} \cdot \vec{k} \sigma_y \otimes \tau_x - \vec{L}_{13} \cdot \vec{k} \sigma_y \otimes \tau_z \quad (9)$$

$\sigma_y, \tau_x, \tau_z$ are all Pauli matrices and two Kronecker product matrices satisfy the anti-commutation relations. Although Eq. (9) is in the form of a massless Dirac equation, $\vec{L}_{14} \cdot \vec{k}$ and $\vec{L}_{13} \cdot \vec{k}$ contain no imaginary parts indicating the zero Berry phase, which is different from the Dirac cone at the corner of the BZ.

### C. Transmission properties

Figure5 shows the numerical simulations of the wave propagating properties in the PC. In panels a1 and a2, we set the operating frequency to $\omega_1 = 0.9409\omega_0$ below the frequency of Dirac point, $\omega_2 = 0.9991\omega_0$ is used in panels b1 and b2 near $\omega_0$. The incident wave is along $\Gamma M$ direction. Figure 5(a1) shows that the outgoing wave preserves the plane wave front, while Fig. 5(b1) shows the Talbot effect.[23] The Talbot effect is a near-field diffraction effect which was first observed in 1836,[24] which means a plane wave is transmitted through a grating or other periodic structure and the resulting wave front propagates in such a way which replicates the structure. According to the field distribution showed in Fig. 5(b1), 5(c1), the wave fronts out of the PC share almost the same shape, only Fig. 5(a1) returns into a plane wave after propagating about $2.3\lambda$ distance. Noted that the widely used effective medium theory is no longer applicable at such high frequency, we cannot expect a plane wave at frequency $\omega_2$ in the PC with $C_{6v}$ symmetry. Here, $\omega_1$ is a threshold frequency to reconstruct the plane wave. For a slight blue shift of frequency $\omega_1$, we can find Talbot effect in our system as shown in Fig. 5(c1).

The defect insensitivity of the Talbot effect in our PC is also investigated and the results are shown in Figs. 5(a2), 5(b2) and 5(c2). Comparing with Dirac point at $K$ point or Dirac-like point at $\Gamma$ point in triangular lattice, the propagation of wave in our PC at Dirac cone frequency is more insensitive to defects. At frequency $\omega_2$, the defect cannot be detected from the transmitted pattern, while it can be easily found at the frequencies of $\omega_1$ and $\omega_3$. According to the field distributions shown in Fig. 5(a2), 5(c2), more than one mode are excited in the PC at frequencies of $\omega_1$ and $\omega_3$. These modes are all attributed by the scattering of the defect, which would provide various scattering wave modes. As the EFC shown in Figs. 3(b1) and 3(b2), it is a circle near frequency $\omega_0$ compared to two big hexagons at the frequencies of $\omega_1$ and $\omega_3$. Considering the incident angle dependence, one wave vector of incident waves would excite two outgoing modes near $\omega_0$, however, such two outgoing modes would share the same Bloch wave vector due to the double degeneracies [Fig. 1]. Furthermore, the existence of defect contributes to wave vectors with various directions. Thus, we can conclude that these two states near $\omega_0$ are insensitive to incidence direction of wave vectors.[25] But, at frequencies of $\omega_1$ and $\omega_3$, two non-degenerated modes are excited corresponding to two different Bloch wave vectors, which are dependent on the incidence angle.

. To check the angle dependent propagation properties in our system, we employ a cylindrical incidence source which can provide various wave vectors [Fig. 6]. Similar to the case with the plane-wave incidence source, Fig. 6(b1) also shows a Talbot effect. The defects in this system cannot be detected at the frequencies near $\omega_0$ [Fig. 6(b2)], which can be regarded as a kind of cloaking.[7,8] The negative refraction also can be realized in Fig. 6(a1).[26,27]

Moreover, the Talbot effect is immune to various types of defects instead of special cases. Figure 7(a) shows the case of a random distribution with several defects, and the transmitted pattern is similar to Fig. 5(b1). With metallic defects ($\rho_2 = 7670 kg/m^3, c_2 = 6010 m/s$, $r = 1m$ in Fig. 7(b) and $r = 1.3m$ in Fig. 7(c)), the scattering of cylinder is well suppressed. To further demonstrate the ability to reduce the scattering cross section, we introduce an air bubble ($r = 1.3m$) which has a strong scattering field in usual underwater acoustic system Fig. 7(d). The scattering of the air bubble is also effectively suppressed.

## IV Summary

In summary, we design a two-dimensional phononic honeycomb lattice to achieve a quadruple-degenerate state at $\Gamma$ point which is constructed by the accidental degeneracy of two double-degenerate states. In the vicinity of the quadruple-degenerate state, there exist double isotropic linear Dirac cones. The linear dispersion induced by the accidental degeneracy is rigorously analyzed by the group representation theory. By using the $\vec{k} \cdot \vec{p}$ method, a 4×4 reduced Hamiltonian is obtained to describe the massless Dirac linear dispersion relation. The Berry phase of such double Dirac cones cancels out due to the absence of the imaginary part. Although there is no flat band in our system, and it neither satisfies long wave approximation nor is regarded as effective zero-index medium, a new kind of novel Talbot effect can still be found in this phononic crystal near the quadruple degenerate point due to the linear and isotropic dispersion, which is insensitive to various types of defects and wave source. Not only that, we also expect *Zitterbewegung* could be observed for such quadruple-degenerate state is associated with Dirac equation.[28, 29] And the enhancement of nonlinear is also prospective for the phase matching effect in our system.[30, 31]

## ACKNOWLEDGEMENTS

The work was jointly supported by the National Basic Research Program of China (Grant No. 2012CB921503 and No. 2013CB632904) and the National Nature Science Foundation of China (Grant No. 1134006). We also acknowledge the support from Academic Program Development of Jiangsu Higher Education (PAPD) and China Postdoctoral Science Foundation (Grant No. 2012M511249 and No. 2013T60521). KAUST Baseline Research fund is also acknowledged.

**References**

[1] Y. B. Zhang, Y. W. Tan, H. L. Stormer, and P. Kim, Nature **438**, 201 (2005).

[2] M. I. Katsnelson, Eur. Phys. J. B **51**, 157 (2006).

[3] M. I. Katsnelson, K. S. Novoselov, and A. K. Geim, Nat. Phys. **2**, 620 (2006).

[4] R. A. Sepkhanov, Y. B. Bazaliy, and C. W. J. Beenakker, Phys. Rev. A **75**, 063813 (2007).

[5] A. H. Castro Neto, F. Guinea, N. M. R. Peres, K. S. Novoselov, and A. K. Geim, Rev. Mod. Phys. **81**, 109 (2009).

[6] D. Torrent and J. Sánchez-Dehesa, Phys. Rev. Lett. **108**, 174301 (2012).

[7] X. Huang, Y. Lai, Z. H. Hang, H. Zheng, and C. T. Chan, Nat. Mater. **10**, 582 (2011).

[8] F. Liu, X. Huang, and C. T. Chan, Appl. Phys. Lett. **100**, 071911 (2012).

[9] F. Liu, Y. Lai, X. Huang, and C. T. Chan, Phys. Rev. B **84**, 224113 (2011).

[10] P. Moitra, Y. Yang, Z. Anderson, I. I. Kravchenko, and D. P. B. J. Valentine4, Nat. Photonics **7**, 791 (2013).

[11] Y. Wu, J. Li, Z.-Q. Zhang, and C. T. Chan, Phys. Rev. B **74**, 085111 (2006).

[12] X. Zhu, B. Liang, W. Kan, X. Zou, and J. Cheng, Phys. Rev. Lett. **106**, 014301 (2011).

[13] S. Zhang, C. Xia, and N. Fang, Phys. Rev. Lett. **106**, 024301 (2011).

[14] K. Sakoda and H. Zhou, Opt. Express **19**, 13899 (2011).

[15] S. G. J. John D. Joannopoulos, Joshua N. Winn, and Robert D. Meade, *Photonic Crystals: Molding the Flow of Light*, 2nd edn.(Princeton University Press, New Jersey, 2008).

[16] J. Mei, Y. Wu, C. T. Chan, and Z. Q. Zhang, Phys. Rev. B **86**, 035141 (2012).

[17] K. Sakoda, Opt. Express **20**, 9925 (2012).

[18] M. Kafesaki and E. N. Economou, Phys. Rev. B **60**, 11993 (1999).

[19] J. Li and C. T. Chan, Phys. Rev. E **70**, 055602 (2004).

[20] Y. Li, Y. Wu, X. Chen, and J. Mei, Opt. Express **21**, 7699 (2013).

[21] G. D. M. S. Dresselhaus, and A. Jorio, *Group Theory: Application to the Physics of Condensed Matter* (Springer-Verlag, Berlin, Heidelberg, 2008).

[22] J. Bolte and S. Keppeler, Ann. Phys. **274**, 125 (1999).

[23] J. R. Leger and G. J. Swanson, Opt. Lett. **15**, 288 (1990).

[24] H. F. Talbot, No. IV, Philos. Mag. 9 (1836).

[25] Q. Cheng, W. X. Jiang, and T. J. Cui, Phys. Rev. Lett. **108**, 213903 (2012).

[26] M.-H. Lu *et al.*, Nat. Mater. **6**, 744 (2007).

[27] J. Christensen and F. Javier Garcia de Abajo, Phys. Rev. Lett. **108**, 124301 (2012).

[28] X. Zhang, Phys. Rev. Lett. **100**, 113903 (2008).

[29] X. Zhang and Z. Liu, Phys. Rev. Lett. **101**, 264303 (2008).

[30] V. Krutyanskiy, I. Kolmychek, E. Gan'shina, T. Murzina, P. Evans, R. Pollard, A. Stashkevich, G. Wurtz, and A. Zayats, Phys. Rev. B **87**, 035116 (2013).

[31] H. Suchowski, K. O'Brien, Z. J. Wong, A. Salandrino, X. Yin, and X. Zhang, Science **342**, 1223 (2013).

**Table caption:**

**Tab. 1** Four states at $\Gamma$ point corresponding to four Bloch bases classified under different symmetry operation of $C_{6V}$ group. $\psi_1$ - $\psi_4$ correspond to field distributions in Figs. 2(a-d), respectively.

**Figures caption:**

Fig. 1(a) Band structures of a 2D honeycomb lattice PC consisting of iron cylinders (radius $r = 0.3710d$) in water. Four linear bands intersect at one point of $\omega_0 = 1.0378 \cdot (2\pi c_1/a)$ in red rectangle region. (b) Band structure with cylindrical radius $r = 0.32d$, the degeneracy is lift.

Fig. 2(a-d) Pressure field distributions of four degenerate Bloch states at $\Gamma$ point as indicated in Fig.1(a) corresponding to $\psi_1$, $\psi_2$, $\psi_3$ and $\psi_4$ from low band to high, respectively. Dark red and dark blue colors denote the positive and negative fields.

Fig. 3(a) The relations of four vectors [Eq.3] in real space calculated by field distribution in Fig. 2. These four real vectors have same length. (b) Dirac dispersion relation. Dots and solid lines represent the simulation results and $\vec{k} \cdot \vec{p}$ method results, respectively.

Fig. 4 Two EFCs corresponding to different bands at the frequencies of $840Hz, 892Hz, 930Hz$. (a1) and (a2) are EFCs of $840Hz$. (b1) and (b2) are EFCs of $892Hz$ near Dirac points. (c1) and (c2) are EFCs of $930Hz$.

Fig. 5 Transmitted patterns with plane incidence source. Operation frequencies are set at (a) $\omega_1$, (b)

$\omega_2$ (c) $\omega_3$. The suffix 1 and 2 represent the cases of intact PC and PC with defect, respectively. (b1) and (c1) exhibit the Talbot effect. (b2) shows the defect-immune property

Fig. 6 Transmitted patterns with plane cylindrical source. Operation frequencies are set at (a) $\omega_1$, (b) $\omega_2$ (c) $\omega_3$. The suffix 1 and 2 represent the cases of intact PC and PC with defect, respectively. (b1) and (c1) exhibit the Talbot effect. (b2) shows the defect-immune property

Fig. 7 Plane wave transmitted pattern with various kinds of defects. (a) Random distributed defects. A metallic cylindrical defect (b) $r = 1m$ and (c) $r = 1.3m$. (d) An air bubble defect ($r = 1.3m$). All of them show a Talbot effects regardless of the type of defects.

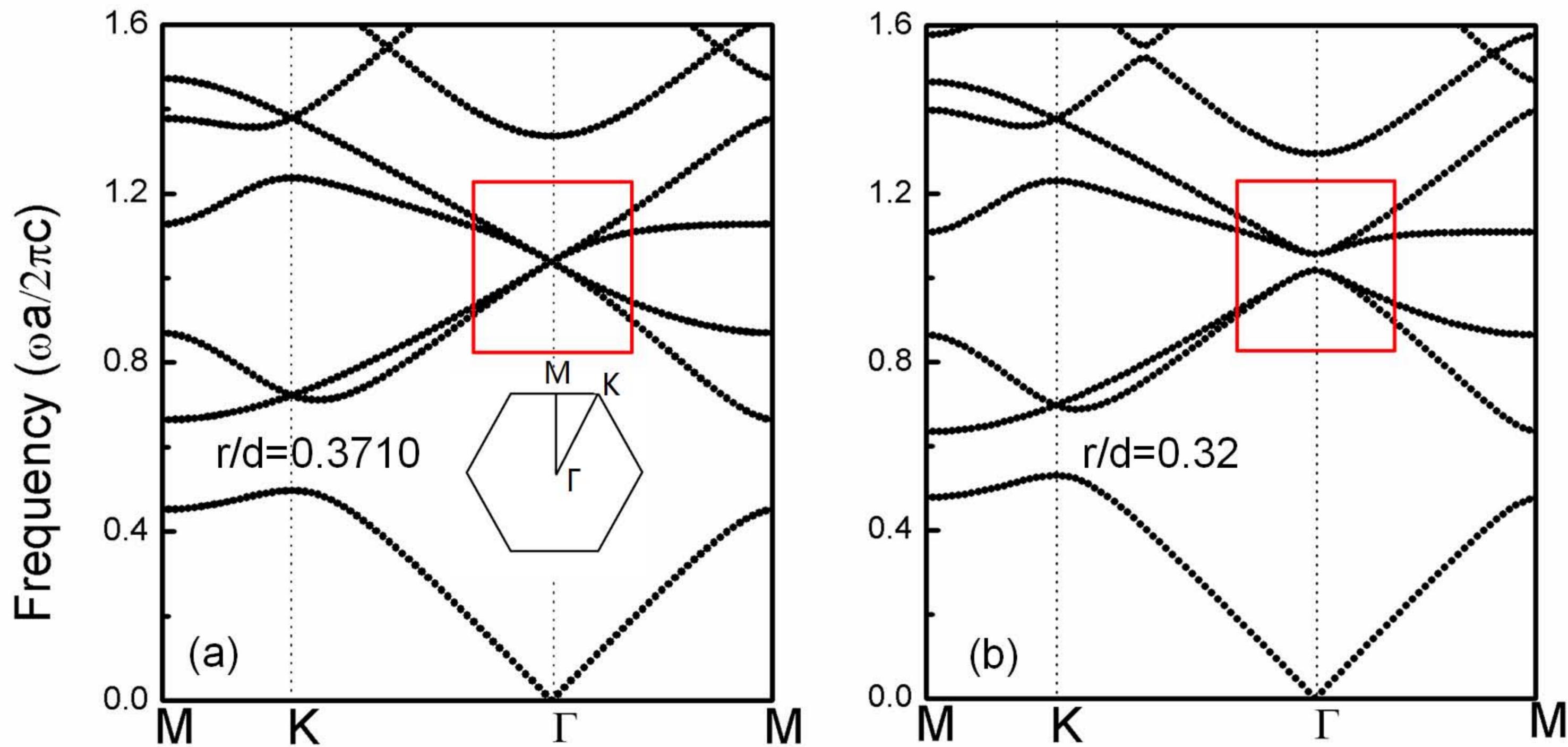

Fig. 1

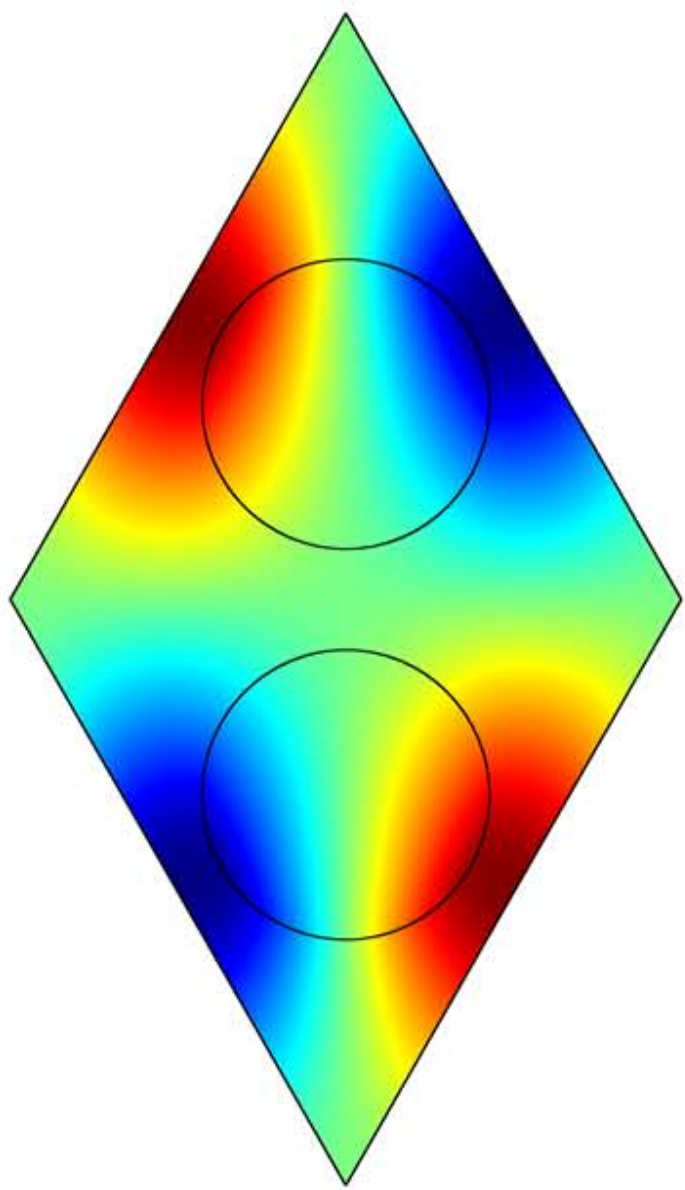

(a)

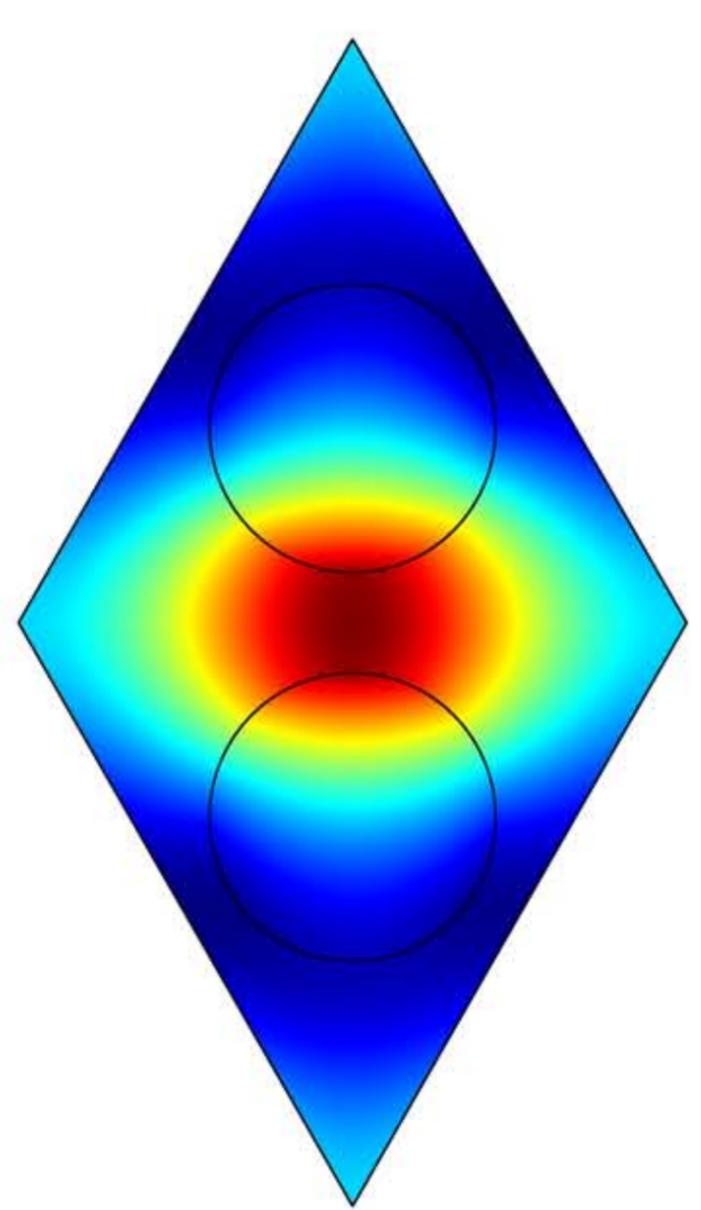

(b)

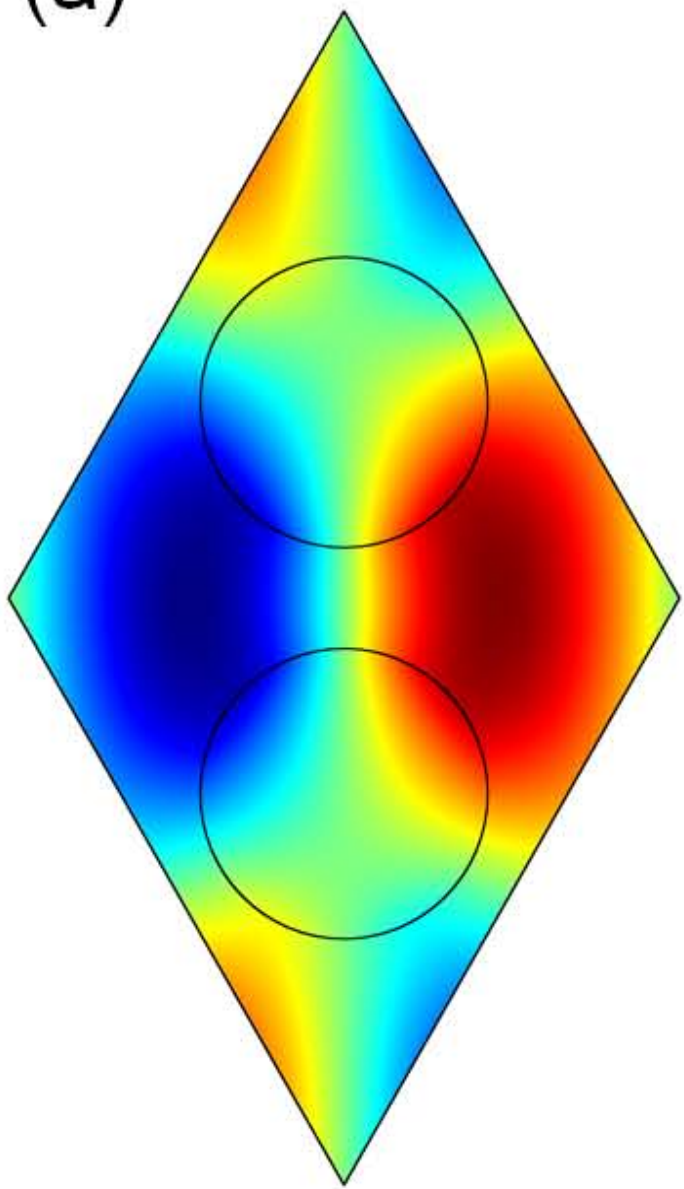

(c)

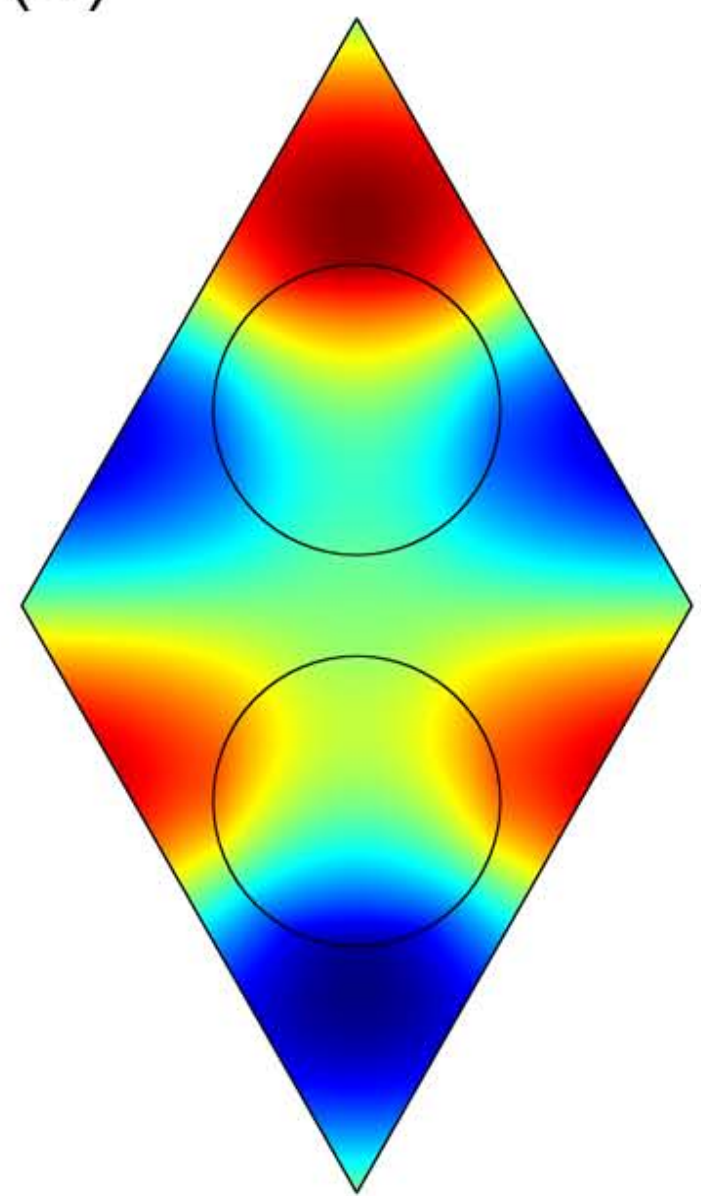

(d)

Fig. 2

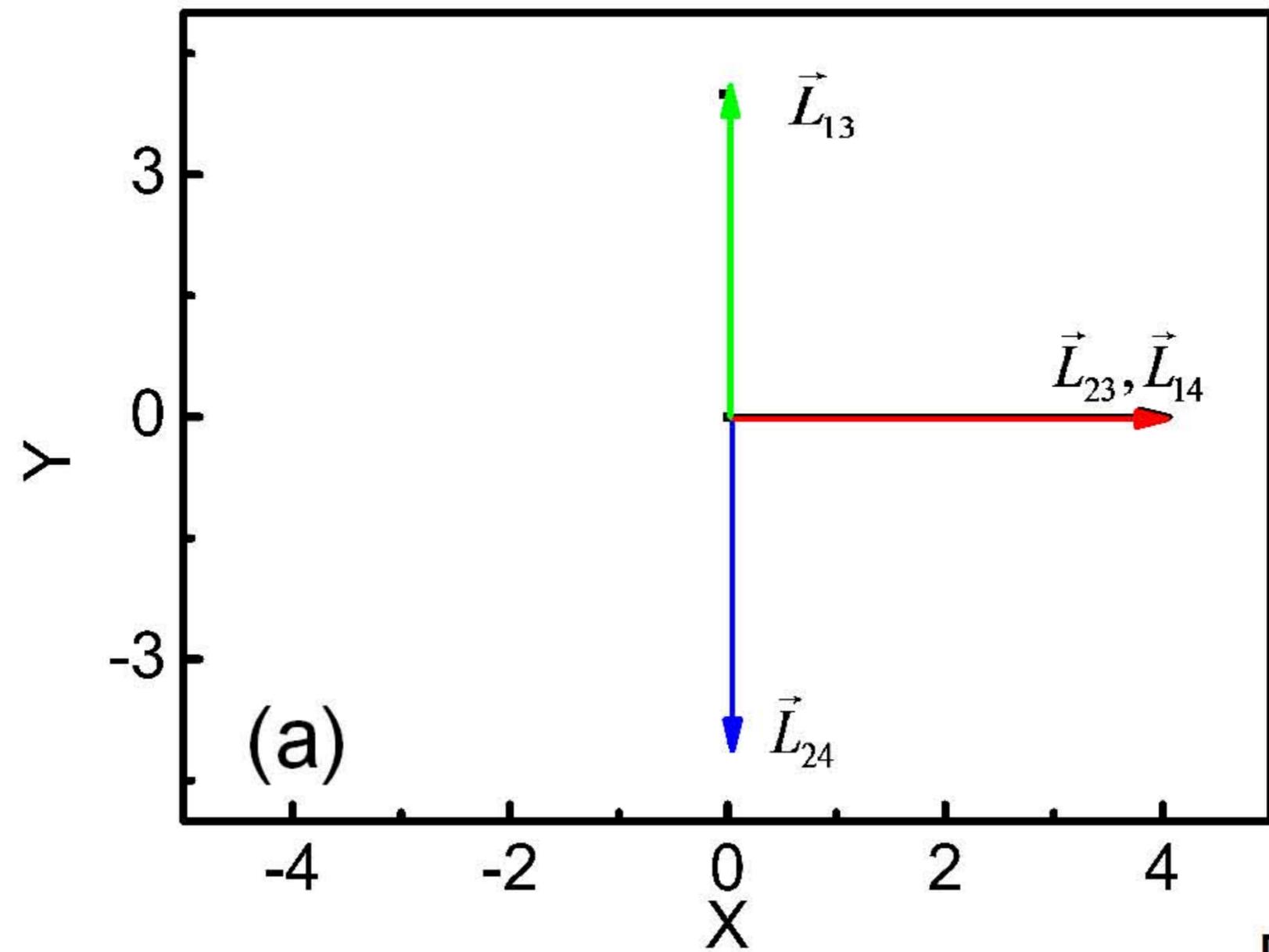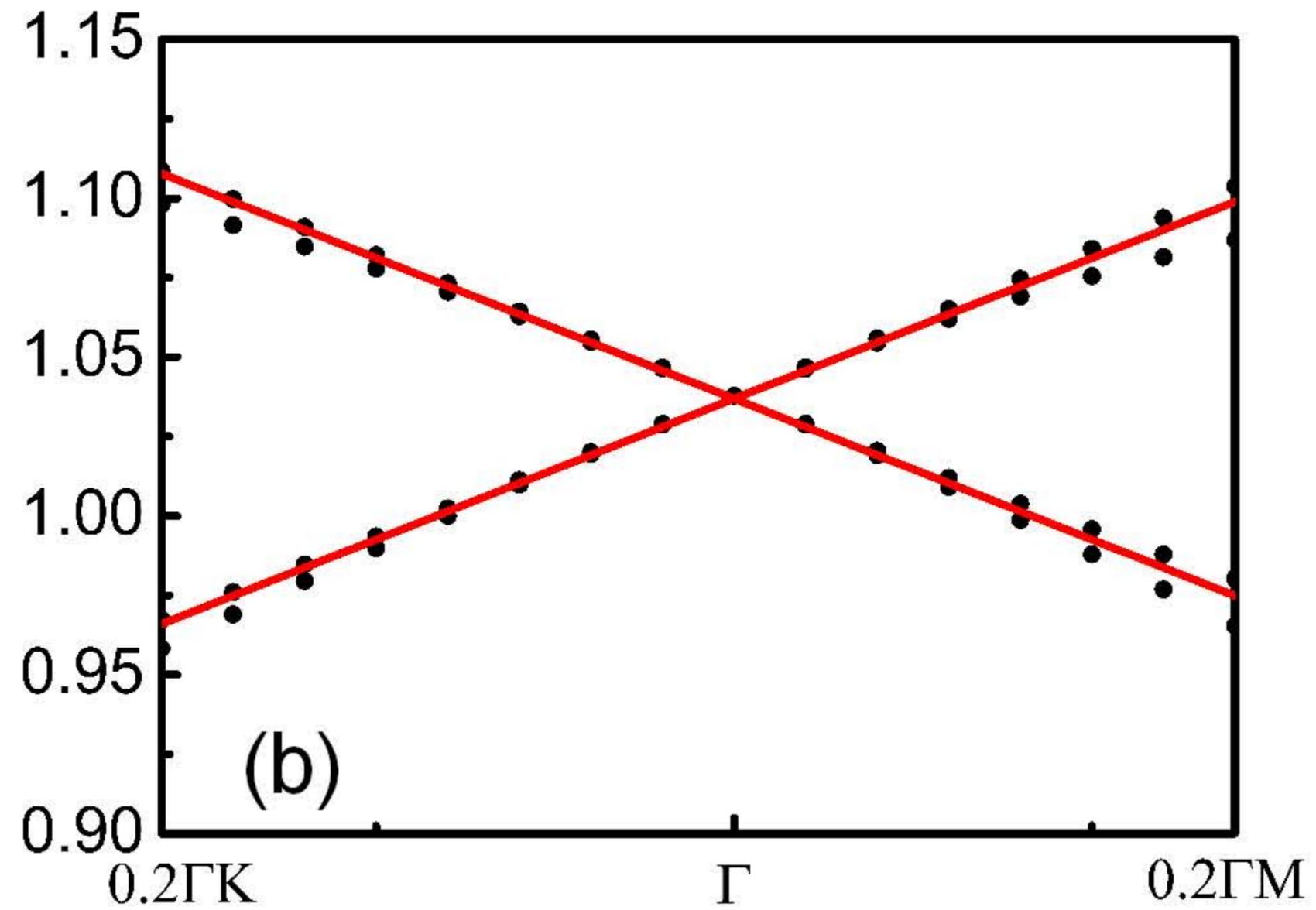

Fig. 3

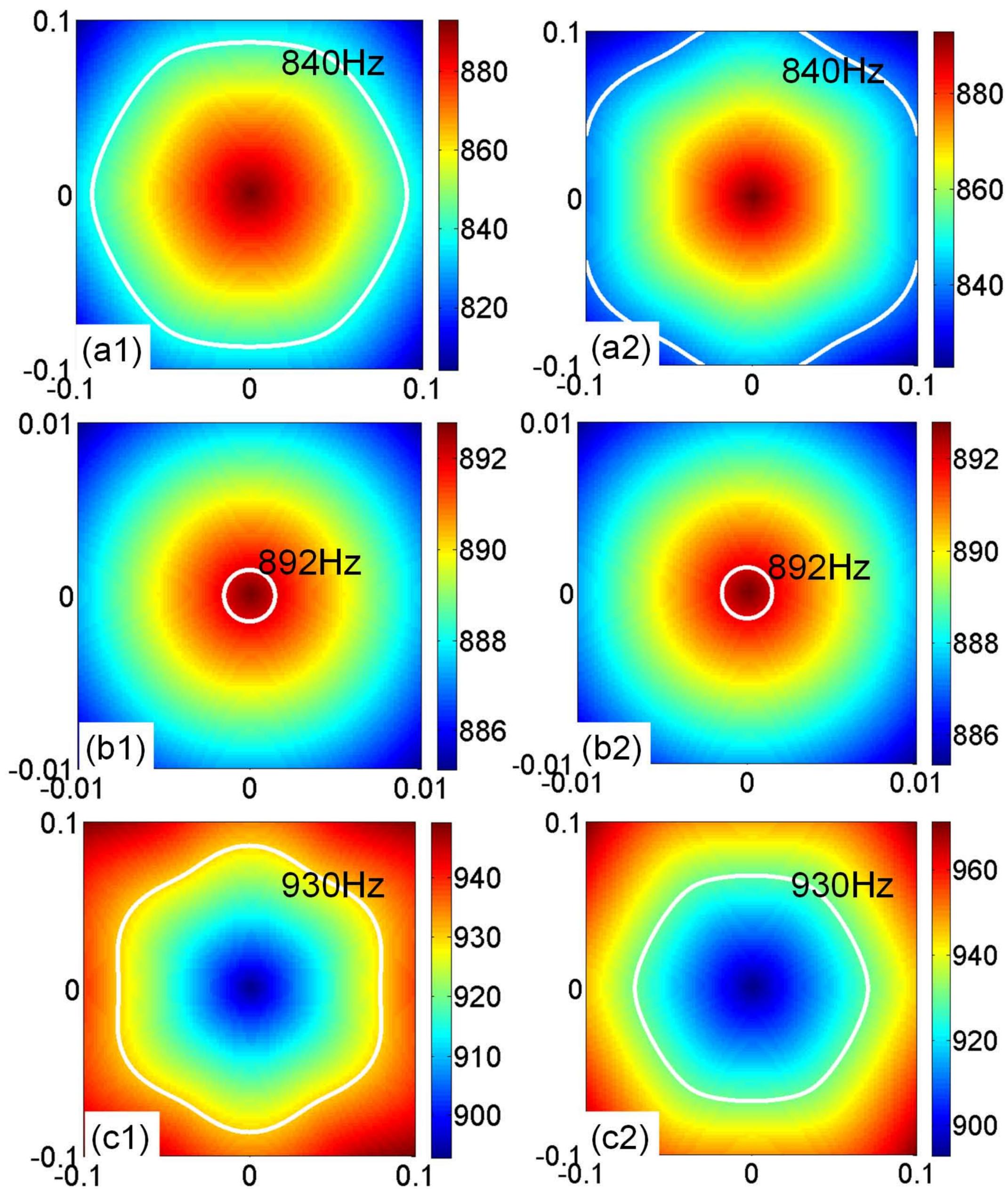

Fig. 4

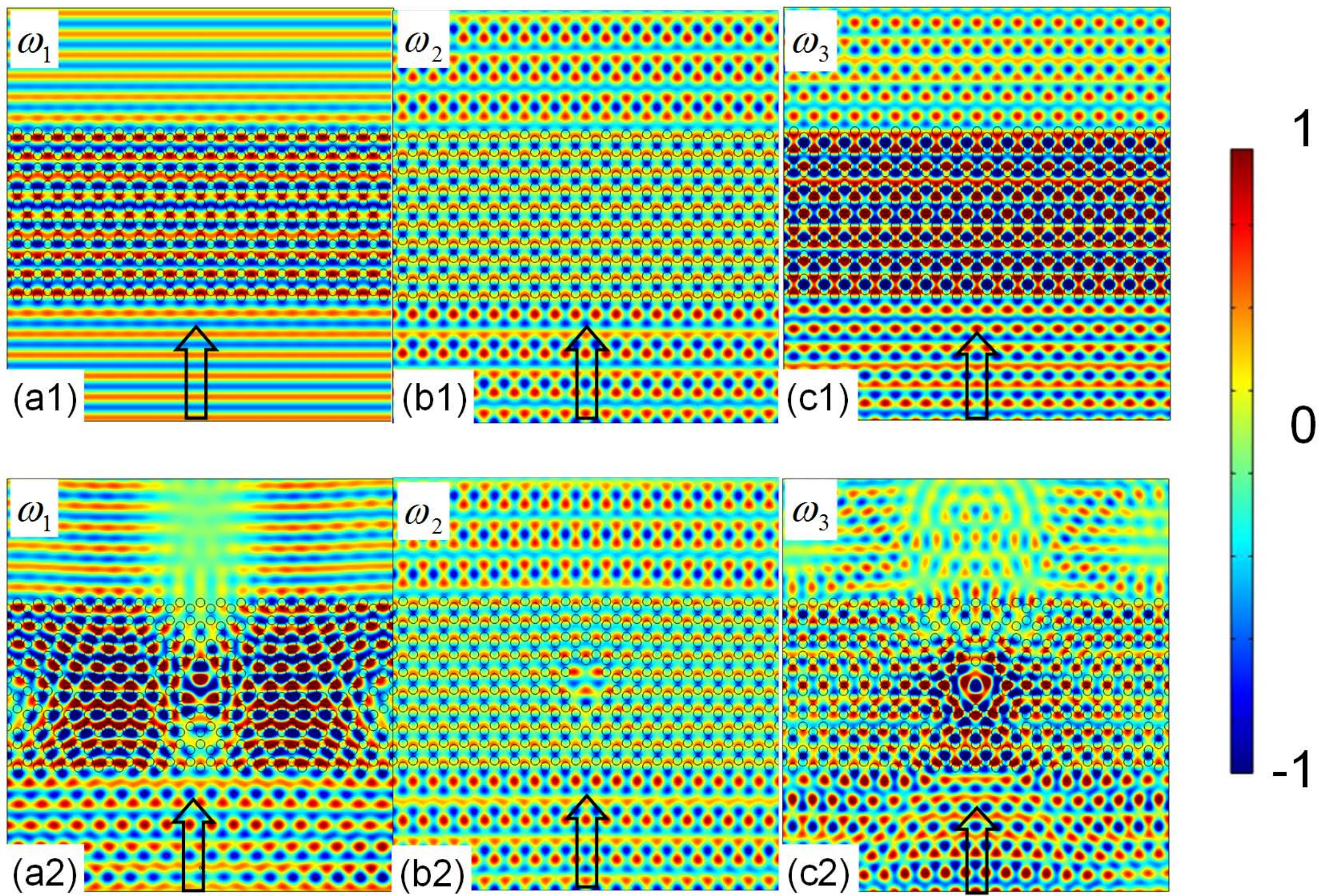

Fig. 5

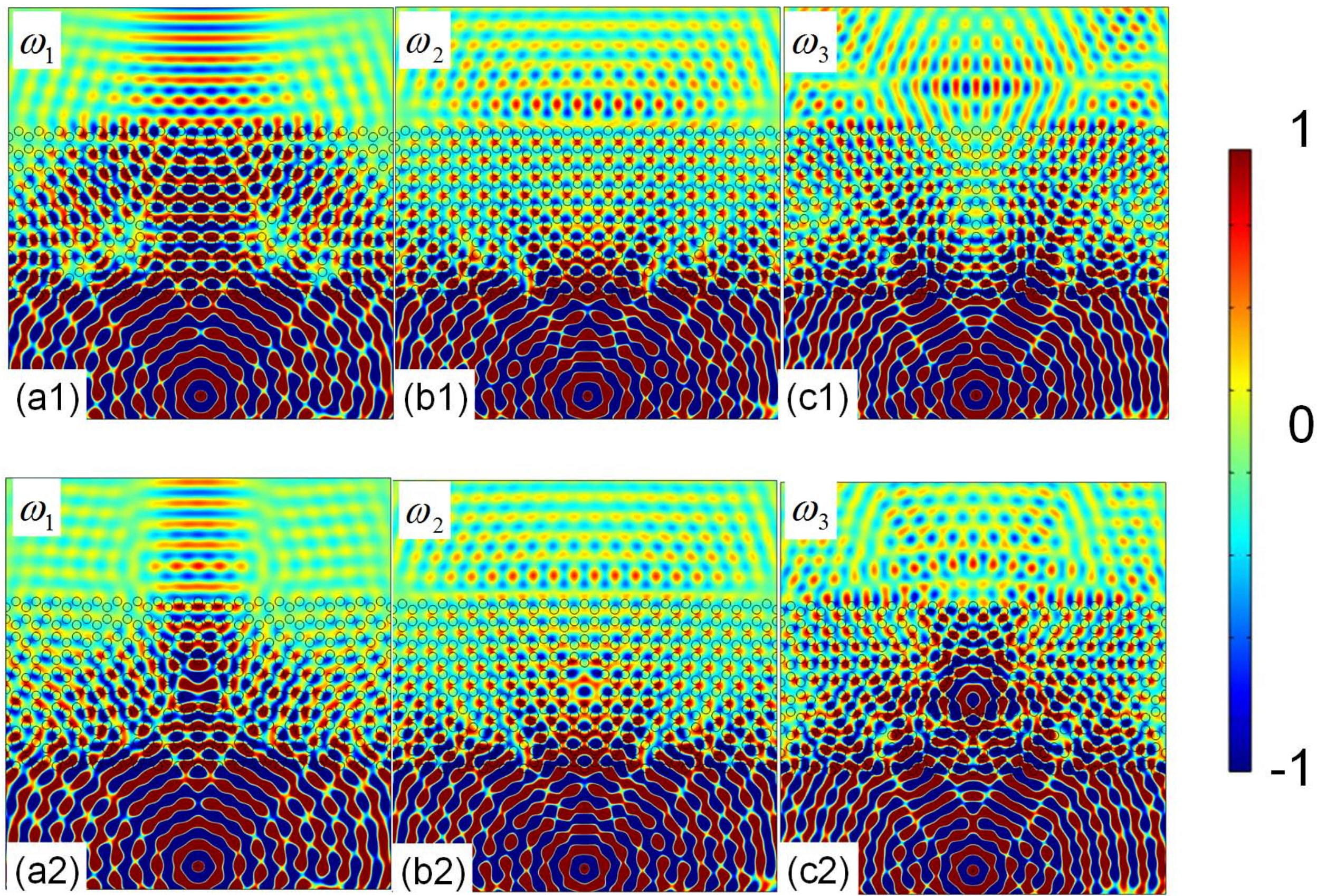

Fig. 6

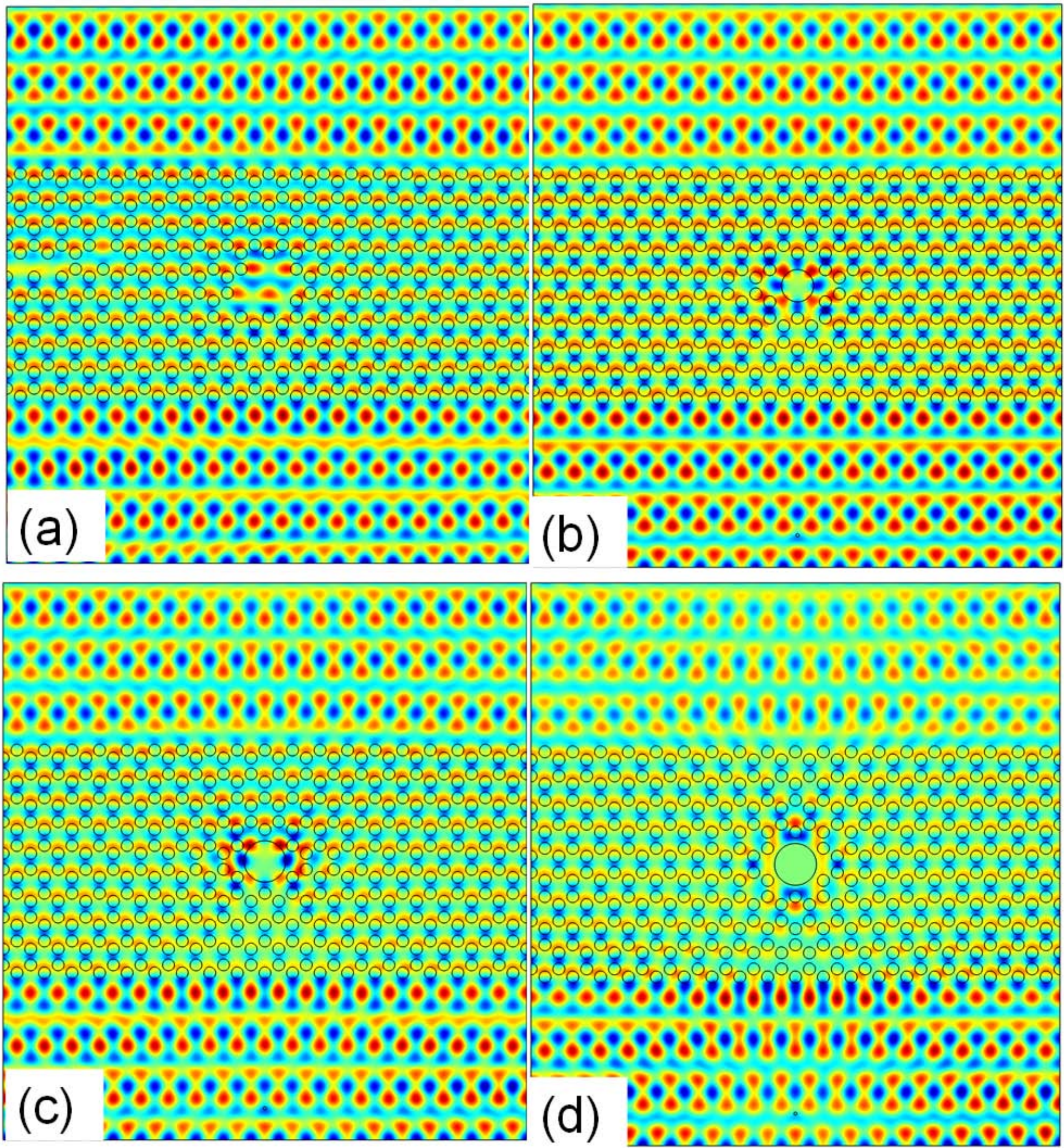

Fig. 7

| $C_{6v}$ | Basis | $\sigma_x$ | $\sigma_y$ |
|---|---|---|---|
| $E_1$ | $\psi_3, x$ | -1 | 1 |
| | $\psi_4, y$ | 1 | -1 |
| $E_2$ | $\psi_1, xy$ | -1 | -1 |
| | $\psi_2, x^2 - y^2$ | 1 | 1 |

Table. 1